\documentstyle[prl,aps,epsf,multicol]{revtex}
\begin{document}
\title{\bf Persistence of a Particle in the Matheron-de Marsily Velocity Field}

\author{Satya N. Majumdar}

\address{Laboratoire de Physique Theorique (UMR 5152 du CNRS), 
Universit\'e Paul Sabatier, 31062 Toulouse Cedex, France. \\}

\date{\today}

\maketitle

\begin{abstract}
We show that the longitudinal position $x(t)$ of a particle in a $(d+1)$-dimensional
layered random velocity field (the Matheron-de Marsily model) can be identified as
a fractional Brownian motion (fBm) characterized by a variable Hurst exponent $H(d)=1-d/4$
for $d<2$ and $H(d)=1/2$ for $d>2$. The fBm becomes marginal at $d=2$. Moreover, using the known 
first-passage properties of fBm we prove analytically that the disorder averaged persistence
(the probability of no zero crossing of the process $x(t)$ upto time $t$) has a power law decay 
for large $t$ with an exponent $\theta=d/4$ for $d<2$ and $\theta=1/2$ for $d\geq 2$ (with
logarithmic correction at $d=2$), results that were earlier derived by Redner based 
on heuristic arguments and supported by numerical simulations (S. Redner, Phys. Rev. E {\bf  
56}, 4967 (1997)). 

\noindent

\end{abstract}

\begin{multicols}{2}

The Matheron-de Marsily (MdM) model,  originally introduced
to study the hydrodynamic dispersion of a tracer particle in porous rocks\cite{MM}, 
provides perhaps the simplest setting where anomalous super-diffusion arises due to 
long range correlations generated dynamically by the process itself. In the MdM model
a single particle diffuses in a $(1+1)$-dimensional layered medium with one transverse and one 
longitudinal direction. The motion of the particle
is purely Brownian along the transverse $y$ direction. In contrast, along the longitudinal
$x$ direction the particle is driven by a drift velocity $v(y)$ that
is a random function of only the transverse coordinate $y$ (see Fig. 1).
Even though the velocities of different layers in the MdM model are uncorrelated,
the motion along the $x$ direction gets correlated in time due to the multiple visits
to the same transverse layer by the particle in a given time $t$. This generates
a typical bias in the $x$ direction giving rise to a super-diffusive longitudinal
transport: $\sqrt {\langle x^2(t)\rangle} \sim t^{3/4}$ for large $t$\cite{BJKP,BG,R1,ZKB}.
\begin{figure}
  \narrowtext\centerline{\epsfxsize\columnwidth \epsfbox{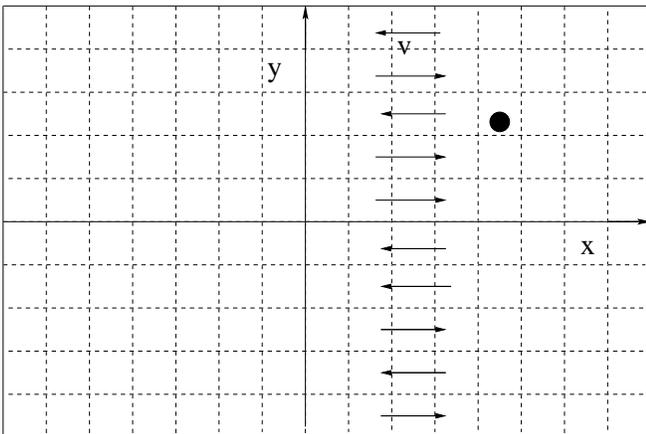}}
\caption{A realization of the MdM model in $(1+1)$ dimensions. The heavy black dot represents the particle.
The velocity field, uniform in a row but fluctuating from row to row, is represented
by the arrows.}
\end{figure}

While the transport properties in the MdM model are by now well understood, the first-passage
properties are much less obvious\cite{R1,R2}. The first-passage property of a particle
diffusing in a random medium is a subject of long standing interest\cite{BG,R3}. 
Quite generally, the first-passage probability $F(t)$ is simply the distribution of the time when a 
particle first crosses a given point in space. The persistence $P(t)$, on the other hand, is 
the probability that the particle does not cross a given point upto time $t$. The two
probabilities are simply related: $F(t)=-dP(t)/dt$\cite{R3}. 
The calculation of the persistence $P(t)$ is
nontrivial even in pure systems without disorder. There have been tremendous theoretical
and experimental activities over the last decade to understand the persistence properties
in nonequilibrium systems without quenched disorder\cite{Review}. Not surprisingly, the persistence 
in disordered systems is even harder to understand. The one dimensional Sinai model\cite{Sinai}, 
where a particle diffuses in a random potential that itself is the trace of a random walk in space,
is perhaps the only model with quenched disorder for which the persistence properties
are understood analytically\cite{Sinai1,Sinai2}. It is, therefore, important to find 
other models with quenched disorder for which the persistence can be
computed analytically. Given its relative simplicity, the MdM model seems to be
a likely candidate for this purpose.

The persistence properties in the MdM model were first studied by Redner\cite{R1,R2}.  
Perhaps the simplest question is: What is the probability $P(t)$ that
the particle does not re-cross the $y$ axis upto time $t$, starting initially at $x(t=0)=0$? This is 
equivalent
to restricting the particle in the semi-infinite geometry ($x>0$) with an absorbing boundary 
condition at $x=0$ and $P(t)$ then represents the survival probability of the particle\cite{R3}. 
The probability $P(t)$, of course, varies from one realization of the velocity field
to another. Redner argued physically, supported by numerical simulations, that the 
disorder averaged survival probability $\overline {P(t)}$ decays as a
power law for large $t$, 
${\overline {P(t)}}\sim t^{-1/4}$\cite{R2}. Extension of this argument to a $(d+1)$-dimensional
generalization of the MdM model (with $d$ transverse directions and one longitudinal
direction) suggests a rather rich asymptotic behavior\cite{R2},
\begin{equation}
{\overline {P(t)}} \sim  \cases
    { t^{-d/4}, \,\quad\quad\,\, &$0<d<2$ \cr
    (\ln t/t)^{1/2}, \,\,\,\, &$d=2$ \cr
    t^{-1/2}, \,\,\quad\quad &$d>2$,}
\label{asymp1}
\end{equation}
indicating that $d=2$ is a `critical' dimension below which the particle survives 
longer (compared to the ordinary diffusion) due to the random velocity field.
The arguments used by Redner in deriving these asymptotic results, though physically
intuitive, are heuristic. Hence, a rigorous derivation would be important and welcome. 

The main result of this Rapid Communication is to show  
that the stochastic process
$x(t)$, representing the longitudinal coordinate of the particle in the $(d+1)$-dimensional MdM 
model, can be identified as 
a fractional Brownian motion (fBm) introduced by Mandelbrot and van Ness \cite{MvN} and is 
characterized by a Hurst exponent $H(d)$ that depends only on  
the dimension $d$. This result shows that the MdM model provides a physical realization of
a fBm with a variable (by varying the dimension $d$ in the MdM model) Hurst exponent. 
This connection to fBm also allows us to prove the asymptotic results in 
Eq. (\ref{asymp1}) by using a known result for the first-passage probability of fBm.

For later purposes it is useful to summarize the definition and the known first-passage
properties of a fBm. A stochastic process $x(t)$ (with zero mean $E\,[ x(t)]=0$) is
called a fBm if its incremental two-time correlation function 
$C(t_1,t_2)=E\,\left[\left(x(t_1)-x(t_2)\right)^2\right]$
is {\em stationary}, i.e., depends only on the difference $|t_1-t_2|$ and moreover, 
grows asymptotically as a power law,
\begin{equation}
C(t_1,t_2) \sim |t_1-t_2|^{2H}, \,\,\,\, |t_1-t_2|>>1. 
\label{hurst1}
\end{equation}
The parameter $0<H<1$ is called the Hurst exponent that characterizes the fBm\cite{MvN} and
$E\,[\dots]$ denotes the expectation value over all realizations of the process. For 
example, the ordinary Brownian motion is a fBm with $H=1/2$. The zero crossing properties of fBm
has been studied before\cite{B,HEM,DW,Krug1} and it is known,
both analytically and numerically, 
that the probability that a fBm does not cross zero upto time $t$ has a power
law decay, $P(t)\sim t^{-\theta}$ for large $t$ with $\theta=1-H$\cite{DW,Krug1}. 
Note that the result $\theta=1-H$ does not require the process
$x(t)$ to be Gaussian and holds for any zero mean process that satisfies Eq. (\ref{hurst1}).
Our strategy, in this paper, would be to first establish that the transverse coordinate $x(t)$ of the 
particle in 
a $(d+1)$- dimensional MdM model indeed satisfies Eq. (\ref{hurst1}) with a Hurst exponent 
$H(d)$ and then, to use the known fBm result $\theta(d)=1-H(d)$  
to prove the results in Eq. (\ref{asymp1}).

In the $(d+1)$-dimensional MdM model considered here, 
the $d>0$ transverse coordinates of the particle, denoted by $y_i(t)$'s 
with $i=1,2\dots, d$, undergo ordinary Brownian motion,
\begin{equation}
{\dot y_i}=\eta_i(t),
\label{yc}
\end{equation}
where $\eta_i$'s are standard zero mean Gaussian white noises with correlators, $\langle 
\eta_i(t)\eta_j(t')\rangle=\delta_{i,j}\delta (t-t')$. The longitudinal coordinate $x(t)$, in contrast, 
is driven by a random drift $v[{\bf y}(t)]$ that depends only on the tranverse coordinates 
${\bf y}(t)=\{y_i(t)\}$,
\begin{equation}
{\dot x}= v[{\bf y} (t)] +\xi(t),
\label{xc}
\end{equation}
where $\xi(t)$ is again a delta correlated zero mean Gaussian white noise and is uncorrelated to the
noises $\eta_i(t)$'s. 
The velocity field $v[{\bf y}]$ is quenched, i.e.,
for a given realization of the function $v$, one first evolves $x(t)$ by Eq. (\ref{xc}) and then
one needs to `disorder average' (denoted by ${\overline {\cdots}}$) over different realizations of the 
random 
function $v$. In the original MdM model in $(1+1)$ dimensions, the random function $v(y)$
is considered to be a Gaussian with zero mean, ${\overline {v(y)}}=0$ and a correlator 
${\overline {v(y_1)v(y_2)}}=\delta(y_1-y_2)$. In the generalized $(d+1)$-dimensional version
considered here, we still choose $v[{\bf y}]$ to be a zero mean Gaussian function, but with a
short-ranged correlator,
\begin{equation}
{\overline {v[{\bf y}_1]v[{\bf y}_2]}}= {1\over {[2\pi d a^2]^{d/2}}} \exp\left[-{ {\left(
{\bf y}_1-{\bf y}_2\right)^2}\over {2 d a^2}}\right].
\label{sr}
\end{equation}
The cut-off $a$ in Eq. (\ref{sr}) represents the correlation length of the velocity field in the
transverse directions. Physically, this mimics the fact that the velocity layers have a finite 
thickness of length $a$. For $d<2$, one can safely take the limit $a\to 0$ and recover 
the delta correlated disorder of the
original MdM model. However, a nonzero cut-off is necessary for $d>2$ in order that the 
MdM model in continuum space is well defined. The physical reason for this nonzero cut-off
is due to the fact that in $d>2$, a {\em point} Brownian particle never meets
another fixed {\em point} in space. The events, where two point particles meet, constitute
a set of measure zero. Hence, to obtain physically meaningful results, it is necessary
to have a finite size for either of the particles in $d>2$. In the present context, it means
that the {\em point} particle in our model will feel the velocity fields in the $d$ transverse 
layers provided the layers have a nonzero thickness.
The choice of a Gaussian function in Eq. (\ref{sr}) just makes the subsequent calculations
easier, but in principle one can choose any short-ranged function in Eq. (\ref{sr}). The
exponent $\theta$ turns out to be independent of this choice.
 
Integrating Eq. (\ref{xc}) we get, 
\begin{equation}
x(t)= \int_0^t \xi(\tau)d \tau + \int_0^t v[{\bf y}(\tau)]d\tau,
\label{xc1}
\end{equation}
where we have assumed $x(0)=0$. In order to relate the process $x(t)$
in Eq. (\ref{xc1}) to a fBm, all we need is to compute the 
expectation value $E\,\left[\left(x(t_1)-x(t_2)\right)^2\right]$
where $E$ denotes an average over all
realizations of $x(t)$, arising from the thermal noises as well as the disorder, i.e.,
$E\,[\cdots]\equiv \langle \overline {\cdots}\rangle = \overline {\langle \cdots\rangle}$.
Clearly, from Eq. (\ref{xc1}) one gets $E\,\left[x(t)\right]=0$. Using
$\langle \xi(t_1)\xi(t_2)\rangle=\delta(t_1-t_2)$, one gets from Eq. (\ref{xc1}),
\begin{equation}
E\,\left[x(t_1)x(t_2)\right]= {\rm {min}}(t_1,t_2)+I(t_1,t_2),
\label{xsol1}
\end{equation} 
where $I(t_1,t_2)$ is given by
\begin{eqnarray}
I(t_1,t_2)&=& \int_0^{t_1}\int_0^{t_2} \langle 
{\overline { v[{\bf y}(\tau_1)]v[{\bf y}(\tau_2)]} }\rangle d\tau_1d\tau_2 \nonumber \\
&=&\alpha \int_0^{t_1}\int_0^{t_2} \langle e^{-\left( {\bf y}(\tau_1)-
{\bf y}(\tau_2)\right)^2/{2da^2}}\rangle d\tau_1 d\tau_2,
\label{di1}
\end{eqnarray}
with $\alpha=[2\pi d a^2]^{-d/2}$. The Gaussian correlator in Eq. (\ref{sr})
has been used in going from the first to 
the second line in Eq. (\ref{di1}).
By integrating Eq. (\ref{yc}), it follows that the vector
\begin{equation}
Y_i=y_i(\tau_1)-y_i(\tau_2)=\int_{\tau_2}^{\tau_1} {\eta}_i (\tau)d\tau 
\label{Y1}
\end{equation}
is a Gaussian random variable since $\eta_i$'s are Gaussian. The vector ${\bf Y}$ 
has zero mean, 
$\langle {\bf Y}\rangle=0$ and a variance $\langle {\bf Y}^2 \rangle= d|\tau_1-\tau_2|$,
computed easily from Eq. (\ref{Y1}) using the property $\langle 
\eta_i(\tau_1)\eta_j(\tau_2)\rangle = \delta_{ij}\delta(\tau_1-\tau_2)$.
The quantity $\langle e^{-{\bf Y}^2/ {2da^2}}\rangle$, that appears inside the integral
on the right hand side of Eq. (\ref{di1}), can then be evaluated 
using the explicit Gaussian form of the distribution
\begin{equation}
P({\bf Y})=[2\pi d|\tau_1-\tau_2|]^{-d/2}\exp\left[{-{\bf Y}^2/{2d|\tau_1-\tau_2|}}\right],
\label{Gauss}
\end{equation} 
and we get
\begin{equation}
I(t_1,t_2)=[2\pi d]^{-d/2}\int_0^{t_1}\int_0^{t_2} {{d\tau_1 d\tau_2}\over {[|\tau_1-\tau_2| + 
a^2]^{d/2}} }.
\label{di2}
\end{equation}
Fortunately, the double integral in Eq. (\ref{di2}) can be explicitly performed. Using this explicit 
form in Eq. (\ref{xsol1}) we get, for all $t_1\geq 0$, $t_2\geq 0$ and $d\neq 2, 4$,
\begin{eqnarray}
E\,\left[x(t_1)x(t_2)\right] &=& A\, {\rm {min}}(t_1,t_2) + B\left[(t_1+a^2)^{\beta} + 
(t_2+a^2)^{\beta} \right . 
\nonumber \\ 
&-& \left . (|t_1-t_2|+a^2)^{\beta}-a^{2\beta}\right], 
\label{xsol2}
\end{eqnarray}          
where $\beta=(4-d)/2$, $A= 1-4 a^{2-d}(2\pi d)^{-d/2}/(2-d)$, and $B=4(2\pi d)^{-d/2}/{(2-d)(4-d)}$.
In the limit $d\to 4$, $B$ diverges but $\beta\to 0$, and the second term in Eq. (\ref{xsol2})
just becomes a logarithm, but stays finite. The other marginal case $d=2$ will be discussed later.
Note that while $B$ is independent of the cut-off $a$, $A$ depends 
explicitly on $a$. 

Putting $t_1=t_2=t$ in Eq. (\ref{xsol2}), we get the results for the variance in the
longitudinal coordinate of the particle for all $t\geq 0$,
\begin{equation}
E\,\left[x^2(t)\right]= A\, t +2 B\left[(t+a^2)^{\beta}-a^{2\beta}\right].
\label{var1}
\end{equation}
This result in Eq. (\ref{var1}) clearly emphasizes the role of the cut-off $a$ and the
critical dimension $d=2$. For $d<2$ or equivalently $\beta=(4-d)/2>1$, it follows from Eq. 
(\ref{var1}) that for large $t$ the second term on the right hand side dominates, giving
rise to a super-diffusion 
$E\,\left[x^2(t)\right]\approx
2B t^{(4-d)/2}$. The cut-off $a$ plays no role for $d<2$ and in fact, one can safely take the
$a\to 0$ limit in Eq. (\ref{var1}) for $d<2$. In contrast, for $d>2$ ($\beta=(4-d)/2<1$), the
first term on the right hand side of Eq. (\ref{var1}) dominates for large $t$ and one gets
ordinary diffusion, $E\,\left[x^2(t)\right]\approx A\, t$ where $A=1-4 a^{2-d}(2\pi 
d)^{-d/2}/(2-d)>0$ depends explicitly on the cut-off $a$. In fact, $A$ diverges in the limit $a\to 
0$. Thus, to get a physically meaningful result, it is necessary to have a finite nonzero cut-off $a$ 
for $d>2$.

The incremental two-time correlation function 
$C(t_1,t_2)=E\,\left[\left(x(t_1)-x(t_2)\right)^2\right]$ can be computed 
from Eq. (\ref{xsol2}) for all $t_1,t_2\geq 0$,
\begin{equation}
C(t_1,t_2)= A\, |t_1-t_2| + 2B \left[(|t_1-t_2|+a^2)^{\beta}- a^{2\beta} \right].
\label{incre1}
\end{equation} 
Thus, in the limit $|t_1-t_2|>>1$, the function $C(t_1,t_2)$ again has two different
asymptotic behaviors depending on whether $d<2$ or $d>2$. In the former case, since 
$\beta=(4-d)/2>1$, one gets, $C(t_1,t_2)\approx 2B |t_1-t_2|^{\beta}$ for large $|t_1-t_2|$. 
As in the case of the variance in Eq. (\ref{var1}), the cut-off $a$ does not play any significant role 
for $d<2$ and one can take safely the limit $a\to 0$ in Eq. (\ref{incre1}).
The incremental correlator is thus stationary and moreover, grows as a power-law 
for large $|t_1-t_2|$ as in Eq. (\ref{hurst1}). This proves that for $d<2$, the 
longitudinal displacement 
$x(t)$ of the particle is indeed a fBm with a Hurst exponent, $H(d)=\beta/2 =1-d/4$.
In the complementary case $d>2$, the first term on the right hand side of Eq. (\ref{incre1})
dominates for large $|t_1-t_2|$ and one gets $C(t_1,t_2)\approx A |t_1-t_2|$ with
$A>0$. Once again, as in Eq. (\ref{var1}), $A$ depends explicitly on the cut-off
$a$ (and diverges as $a\to 0$) and it is necessary to retain a nonzero cut-off $a>0$ for $d>2$.
With a finite cut-off, $C(t_1,t_2)$ again has the fBm form as in Eq. (\ref{hurst1}) with
a Hurst exponent $H(d)=1/2$ for all $d>2$. This indicates that $x(t)$ asymptotically becomes an 
ordinary
Brownian motion (or fBm with $H=1/2$) for all $d>2$. This shows that for all $d\neq 2$,
the longitudinal coordinate $x(t)$ of the particle is a fBm with a Hurst exponent, 
\begin{equation}
H(d) =\cases
{ 1-{d\over {4}}, \,\,\,\, &$d<2$ \cr
     {1\over {2}}, \,\,\,\, &$d>2$.\cr} 
\label{hurst2}
\end{equation}   

It then follows immediately from the known first-passage property of fBm mentioned earlier
that the probability, that the process $x(t)$ does not cross zero upto time $t$, decays as
a power law, ${\overline {P(t)}}\sim t^{-\theta}$ with $\theta=1-H$. Using the results
for $H$ in Eq. (\ref{hurst2}), we get $\theta=d/4$ for $0<d<2$ and $\theta=1/2$ for $d>2$,
thus proving the results in Eq. (\ref{asymp1}) for $d\neq 2$.  

In the marginal case $d=2$, one can again evaluate the integral $I(t_1,t_2)$ in Eq. 
(\ref{di2}) explicitly. Substitution of this form in Eq. (\ref{xsol1}) gives for all $t_1, 
t_2 \geq 0$,
\begin{eqnarray}
E\,\left[x(t_1)x(t_2)\right] &=& A' \, {\rm {min}}(t_1,t_2) + B'\left[(t_1+a^2)\ln 
(t_1+a^2)  \right . \nonumber \\
&+& \left . (t_2+a^2)\ln (t_2+a^2)-a^2\ln (a^2) \right. \nonumber \\ 
&-& \left . (|t_1-t_2|+a^2)\ln (|t_1-t_2|+a^2) \right],
\label{corrd2}
\end{eqnarray}
where $A'= 1- \ln(a^2 e)/{2\pi}$ and $B'=1/{4\pi}$. For $t_1=t_2=t$, one gets from Eq. (\ref{corrd2})
the exact result for the variance for all $t$,
\begin{equation}
E\,\left[x^2(t)\right]= A'\, t + 2B'\left[(t+a^2)\ln (t+a^2)-a^2\ln(a^2)\right].
\label{var2}
\end{equation}
Thus, asymptotically for large $t$, the longitudinal motion for $d=2$ is super-diffusive, but only 
marginally 
$E\,\left[x^2(t)\right]\approx 2B' t\ln (t)$. One can easily compute the incremental two-time
correlator $C(t_1,t_2)=E\,\left[\left(x(t_1)-x(t_2)\right)^2\right]$ from Eq. (\ref{corrd2}),
\begin{eqnarray}
C(t_1,t_2)&=& A'\, |t_1-t_2| + 2B'\left[(|t_1-t_2|+a^2)\times \right . \nonumber \\
&\times & \left . \ln 
(|t_1-t_2|+a^2)-a^2\ln(a^2)\right].
\label{incre2}
\end{eqnarray}
Evidently, this correlator is stationary and for $|t_1-t_2|>>1$, it has
a power-law (with logarithmic correction) dependence on its argument as in Eq. (\ref{hurst1})
with $H=1/2$. Thus for $d=2$, the longitudinal motion $x(t)$ of the particle is
a marginal fBm with $H=1/2$. The analytical arguments leading to the
persistence exponent $\theta=1-H$ for fBm\cite{DW,Krug1} can be easily modified
to deal with the marginal case. This gives, ${\overline {P}}(t)\sim \sqrt{\ln (t)/t}$ for large $t$,
thus recovering the result in Eq. (\ref{asymp1}) for $d=2$. 

In summary, we were able to show exactly that the longitudinal position $x(t)$ of a
particle in a $(d+1)$-dimensional MdM model can be represented as a fBm with a
Hurst exponent $H(d)$ that varies with the dimension $d$ of the MdM model.
Moreover, this exact connection allows us to use the known first-passage results for the fBm 
and derive analytically the asymptotic results for
the disorder averaged survival probability in the MdM model that were known before
only via heuristic arguments and numerical simulations\cite{R2}. This method of connecting
to fBm and the subsequent usage of the first-passage results for the fBm was first used to calculate 
exactly the first-return exponent of a fluctuating Gaussian interface to a fixed stationary 
configuration\cite{Krug1}, a result which was subsequently verified in experiments
on step edge fluctuations in Al/Si(111) surface\cite{Exp1}. In this paper, we have shown that
the same connection is useful in calculating the first-passage properties in a model
with quenched disorder.
Given that this technique is useful in two rather different problems, one 
expects that this method perhaps has much wider future applications. 

This paper has dealt with the persistence of a {\em single} particle in the MdM 
velocity field.
It would be interesting to extend this study to a system of {\em interacting} particles. A simple
physical example of such a system is a single polymer chain within the Rouse model 
where the beads on the chain are connected by harmonic springs\cite{Rouse}. The transport
properties of a Rouse chain in presence of an MdM velocity field have been well 
studied in the literature\cite{OB,WL,JOB}. 
The techniques presented in this paper may be extended
to study the persistence properties in a Rouse chain advected by an
MdM velociy field\cite{MDC}.  

It is a pleasure to thank S. Redner, P.L. Krapivsky and A. Comtet for useful conversations.

\end{multicols}

\end{document}